\documentclass[preprintnumbers,amsmath,amssymbm,preprint]{revtex4}
\usepackage{epsfig}
\usepackage{graphicx}

\begin{document}
\title{Quantitative description of cognitive fatigue in repetitive monotonous tasks}
\author{Shahar Hod}
\affiliation{The Ruppin Academic Center, Emeq Hefer 40250, Israel}
\affiliation{ }
\affiliation{The Hadassah Institute, Jerusalem 91010, Israel}
\date{\today}

\begin{abstract}
\ \ \ There is strong qualitative empirical evidence in the scientific literature that, due to cognitive fatigue, 
workers performing repetitive and monotonous tasks are characterized by a gradual deterioration in their performance abilities 
as the time-on-task increases, a phenomenon known as the vigilance decrement. 
Using a time-dependent Sisyphus random climb model, 
we provide a quantitative description of this intriguing phenomenon. 
In particular, we use analytical techniques in order to determine the success probability function $S(t;{\cal N})$ 
of Sisyphus workers, the time-dependent fraction of workers who succeed, 
after making $t$ repetitive operations or less, to complete their task by making ${\cal N}$ successful 
operations in a row without a single fault in between. 
It is explicitly shown that the functional behavior of the increasing-in-time one-operation tumble probability $1-s(t)$ of exhausted Sisyphus 
workers may have a dramatic effect on the 
probability of the workers to achieve their ultimate goal in repetitive monotonous processes. 
In particular, we prove that the Sisyphus random climb model with 
the inverse power law functional behavior $s(t)\sim t^{-1/{\cal N}}$ of the one-operation success probability 
marks the boundary between 
Sisyphus workers whose success functions $S[t;s(t),{\cal N}]$ approach $1$ asymptotically in time 
(implying that all the workers eventually complete their task) and 
Sisyphus workers whose success functions approach an asymptotic value 
which is less than $1$, in which case some of the exhausted Sisyphus workers
never complete their task successfully. 
\end{abstract}
\bigskip
\maketitle

\section{Introduction}
The gradual decrease in vigilance in sustained attention tasks frequently 
appears in the scientific literature as an example of cognitive fatigue \cite{Cap1,Cap2,Cap3,Cap4,Cap5}. 
Interestingly, there is strong empirical evidence that the gradual decline in the performance abilities of 
individuals is even more likely to occur in repetitive tasks \cite{Mesh1,Mesh2,Mesh3}. 
In particular, it is well established experimentally \cite{NW1,NW2,NW3} that sustained attention tasks usually put 
pressure on the cognitive system, leading to performance degradation of most people over time \cite{NW1}. 
This intriguing phenomenon is referred to in the scientific literature as the vigilance decrement, 
defined as the time-dependent probability to perform critical errors as the time-on-task increases.

Resource Theory is the dominant paradigm for explaining the observed dramatic effects of the time-on-task and 
fatigue on the performance abilities of individuals in repetitive sustained attention tasks \cite{NW1,NW2,NW3}. 
According to this theory, sustained attention tasks put high cognitive demands on attention resources 
that gradually produce the vigilance decrement which is known to characterize many monotonous repetitive tasks. 
In particular, there is strong empirical 
evidence in the literature (see \cite{NW1,NW2,NW3} and references therein) that, although many vigilance tasks are quite simple, 
the difficulty in performing these tasks is related to the fact that they demand a continuous long-term focus 
on repetitive operations.  

The goal of the present study is to provide an analytical {\it quantitative} description of the empirically 
observed \cite{Cap1,Cap2,Cap3,Cap4,Cap5,Mesh1,Mesh2,Mesh3,NW1,NW2,NW3} performance degradation over time 
of most individuals in {\it repetitive} tasks. 
In particular, we suggest to model the effects of the time-on-task and the cumulative fatigue 
on the performance abilities of humans in repetitive and monotonous tasks 
using a time-dependent Sisyphus random climb model \cite{RS1,RS2,RS3,RS4,RS5,RS6,Sisy,Hodanp,Hodbia,Hodmpt}. 

In our model the Sisyphus worker is required to perform the same monotonous operation over and over again with 
a time-dependent probability $s(t)$ to succeed in her $t$-th operation and a complementary 
probability $1-s(t)$ to fail and fall back to her starting point, 
resulting in the need to restart the Sisyphean `climbing' process again. 
The ultimate goal (the target) of the Sisyphus worker is achieved if she makes ${\cal N}$ 
successful repetitive operations in a row without a single stumble in between. 

The repetitive task (the `climbing' process) of the Sisyphus worker may be, for example: (1) To memorize ${\cal N}$ random 
numbers, with the need to restart the process all over again (with new random numbers) 
each time an error occurs \cite{Dig1,Dig2}, 
(2) to hit ${\cal N}$ three pointers in a row, with the need to reset the count with each miss of the player \cite{3points}, 
(3) to succeed ${\cal N}$ times in a row in the Stroop test \cite{Stro1,Stro2}, or 
(4) to construct a ${\cal N}$ floor house of cards which may collapse each time 
a new card is added by the worker.

Following the empirical evidence that has been mounting in the scientific literature \cite{Cap1,Cap2,Cap3,Cap4,Cap5,Mesh1,Mesh2,Mesh3,NW1,NW2,NW3} for the 
gradual time-dependent decrease in the performance abilities of most workers in repetitive tasks, 
we shall take into account that the stumble probability $1-s(t)$ of the Sisyphus worker in each repetitive operation 
increases over time. 

The dynamics of the Sisyphus workers (Sisyphus random climbers) is 
characterized by a time-dependent success probability function $S[t;s(t),{\cal N}]$ which determines 
the probability for a random walker to reach her final target. 
In other words, the success function $S[t;s(t),{\cal N}]$ measures the fraction of Sisyphus workers who succeeded, 
after making $t$ repetitive operations or less, 
to make ${\cal N}$ successful operations in a row 
without a single stumble in between. 

Our main goal is to determine, using {\it analytical} techniques, the time-dependent functional behavior of 
the success probability function $S[t;s(t),{\cal N}]$ 
of Sisyphus random workers to reach their final target in real-life situations in which 
the one-operation probability $1-s(t)$ of an exhausted worker to make a mistake (to stumble and fall 
back to her starting point) evolves in time. 

\section{Description of the time-dependent Sisyphus random climb model}
We shall use analytical techniques in order to analyze the functional behavior 
of the success function $S[t;s(t),{\cal N}]$ 
which describes the time-dependent fraction of Sisyphus workers (Sisyphus climbers) who 
have reached their target (completed their task) by making ${\cal N}$ successful repetitive operations
in a row without a single stumble.
 
A Sisyphus climber, who starts her
dynamics at the bottom $h(t=0)=0$ of a `ladder' a distance of ${\cal N}$ successful jumps 
from her target, is characterized by a
time-dependent probability $s(t)$ to take an upward step of length $\Delta h$ towards the top of the ladder (towards 
her final target) and a complementary probability
$1-s(t)$ to stumble and fall back all the way to the bottom of the ladder. 
The space-time evolution of the Sisyphus workers is therefore determined 
by the jumping rule
\begin{equation}\label{Eq1}
h(t+\Delta t)=
\begin{cases}
h(t)+\Delta h& \ \text{with probability}\ \ s(t)\ ;
\\ 0 & \ \text{with probability}\ \ 1-s(t)\  .
\end{cases}
\end{equation}
In order to reach her final target (the top of the ladder), the Sisyphus random climber must 
make ${\cal N}$ jumps in a row in the upward direction without a single stumble in between. 

The evolution equation (\ref{Eq1}) is supplemented by the initial condition
\begin{equation}\label{Eq2}
N_{\text{tot}}(t)=1 \ \ \ \ \text{for}\ \ \ \ t<{\cal N}\cdot\Delta t\ ,
\end{equation}
where 
\begin{equation}\label{Eq3}
N_{\text{tot}}(t)=\sum_{k=0}^{k={\cal N}-1}N_k(t)\
\end{equation}
is the total number of Sisyphus climbers [with the normalization condition 
$N_{\text{tot}}(t=0)=1$] who
have not reached their target after making $t/\Delta t$ steps ($t/\Delta t$ repetitive operations). 
Here $N_k(t)$ is the time-dependent number of Sisyphus climbers who are located at the $k$th level 
(with $0\leq k<{\cal N}$) at time $t$.

\section{The characteristic recurrence relation of the time-dependent success function $S[t;s(t),{\cal N}]$}
In the present section we shall derive an analytical master equation for the time evolution 
of the success probability function $S[t;s(t),{\cal N}]$ which characterizes 
the Sisyphus random climb model (\ref{Eq1}) with generic (time-dependent) jumping probabilities.
 
To this end, it is convenient to define the time-dependent failure probability function
\begin{equation}\label{Eq4}
F[t;s(t),{\cal N}]\equiv 1-S[t;s(t),{\cal N}]\  ,
\end{equation}
which is the fraction of workers who have not completed 
their task (who have not reached the top $h_{\text{top}}={\cal N}\cdot\Delta h$ of the ladder) at time $t$ 
[that is, $F(t)$ is the fraction of Sisyphus random climbers 
who have not succeeded to make ${\cal N}$ successful operations in a row after making a total of $t/\Delta t$ 
repetitive operations]. 
Taking cognizance of the normalization condition (\ref{Eq2}), one finds the identity
\begin{equation}\label{Eq5}
F[t;s(t),{\cal N}]\equiv N_{\text{tot}}(t)\  .
\end{equation}

In order to derive the recurrence relation that governs the temporal evolution of 
the failure probability function $F[t;s(t),{\cal N}]$ of the Sisyphus workers, we shall first 
note that the characteristic jumping rule (\ref{Eq1}) of the Sisyphus climbers implies that the number 
\begin{equation}\label{Eq6}
r(t)\equiv N_{\text{tot}}(t-\Delta t)-N_{\text{tot}}(t)\
\end{equation}
of climbers who reach their target (the top of the ladder at $h_{\text{top}}={\cal N}\cdot\Delta h$) 
after making exactly $t/\Delta t$ jumps is given by the simple relation
\begin{equation}\label{Eq7}
r(t)=s(t-\Delta t)\cdot N_{{\cal N}-1}(t-\Delta t)=s(t-\Delta t)\cdot s(t-2\Delta t)\cdot
N_{{\cal N}-2}(t-2\Delta t)=\cdots =Q\cdot N_0(t-{\cal N}\cdot\Delta t)\  ,
\end{equation}
where the time-dependent jumping probability function $Q(t)$ is defined by the ${\cal N}$-term product
\begin{equation}\label{Eq8}
Q(t;{\cal N})\equiv \prod_{k=1}^{k={\cal N}} s(t-k\cdot\Delta t)\  .
\end{equation}

It is worth emphasizing that the analytically derived relation (\ref{Eq7}) reflects the fact that 
a Sisyphus random climber who is located at the bottom $h=0$ of the ladder at time $t-{\cal N}\cdot\Delta t$ must jump
${\cal N}$ times in a row in the upward direction (without a single stumble in between) in order to reach the top of the ladder 
$h_{\text{top}}={\cal N}\cdot\Delta h$ exactly at time $t$ 
[note that the jumping rule (\ref{Eq1}) implies that an upward step 
towards the top of the ladder is characterized by the time-dependent one-operation success function $s(t)$].

In addition, the jumping rule (\ref{Eq1}), which characterizes the dynamics of the Sisyphus random climbers, 
implies that, at each time step, a Sisyphus climber has a (time-dependent) probability $1-s(t)$ 
to stumble and fall back all the way to the bottom of the ladder. 
Using this fact, one can express the time-dependent number $N_0(t)$ 
of Sisyphus climbers who are located at the bottom of the ladder at time $t$ in the compact form
\begin{equation}\label{Eq9}
N_0(t)=[1-s(t-\Delta t)]\cdot N_{\text{tot}}(t-\Delta t)\  ,
\end{equation}
which implies \cite{Notetng}
\begin{equation}\label{Eq10}
N_0(t-{\cal N}\cdot\Delta t)=[1-s(t-{\cal N}\cdot\Delta t-\Delta t)]
\cdot N_{\text{tot}}(t-{\cal N}\cdot\Delta t-\Delta t)\  .
\end{equation}

From Eqs. (\ref{Eq5}), (\ref{Eq6}), (\ref{Eq7}), and (\ref{Eq10}) one obtains the characteristic master equation \cite{Notetnnn} 
\begin{equation}\label{Eq11}
F(t)=F(t-\Delta t)-Q(t)\cdot[1-s(t-{\cal N}\cdot\Delta t-\Delta t)]\cdot F(t-{\cal N}\cdot\Delta t-\Delta t)\ 
\end{equation}
for the time-evolution of the failure probability function $F[t;s(t),{\cal N}]$ which characterizes the dynamics of 
Sisyphus random climbers with time-dependent jumping probabilities. The success function $S[t;s(t),{\cal N}]$ 
of the physical system, which is defined as the time-dependent fraction of Sisyphus random climbers who 
have reached the top of the ladder (that is, the fraction of climbers who have succeeded in making ${\cal N}$ jumps 
in a row in the upward direction) is given by the complementary functional relation (\ref{Eq4}). 

\section{Success probabilities of exhausted Sisyphus random workers}
In the present section we shall analyze the temporal functional behavior of the success probability fucntion 
$S[t;s(t),{\cal N}]$ that characterizes the dynamics of exhausted Sisyphus workers 
with time-dependent one-operation success probabilities. 

In order to facilitate a fully analytical treatment of the system we shall assume that the one-operation success
probability of the exhausted Sisyphus climbers is characterized by the relation \cite{NoteqQ}
\begin{equation}\label{Eq12}
s(t\geq{\cal N})\ll1\  ,
\end{equation}
in which case the master equation (\ref{Eq11}), which determines the temporal evolution of the failure probability 
function in the Sisyphus random climb model (\ref{Eq1}), can be written in the form
\begin{equation}\label{Eq13}
F(t)=F(t-\Delta t)-Q(t)\cdot F(t-{\cal N}\cdot\Delta t-\Delta t)\  .
\end{equation}

In addition, assuming the characteristic relation (\ref{Eq12}) of the exhausted Sisyphus random climbers, 
the failure probability function can be expanded in the continuum limit in the form
\begin{equation}\label{Eq14}
F(t-k\cdot\Delta t)=F(t)-{{dF}\over{dt}}\cdot(k\Delta t)+O\big[{{d^2F}\over{dt^2}}\cdot(k\Delta t)^2\big]\  .
\end{equation}
Substituting Eq. (\ref{Eq14}) into Eq. (\ref{Eq13}), one obtains the 
remarkably compact differential equation \cite{Notecl}
\begin{equation}\label{Eq15}
{{dF}\over{dt}}=-Q(t)\cdot F\  ,
\end{equation}
which yields the functional expression 
\begin{equation}\label{Eq16}
F(t)=F_0\cdot\exp\Big[-\int_{{\cal N}}^{t} Q(t')dt'\Big]\ \ \ \ \text{for}\ \ \ \ t\geq{\cal N}\
\end{equation}
for the time-dependent failure probability function of the system, 
where the pre-factor $F_0$ in the integral relation (\ref{Eq16}) can be determined from 
the simple probability relation $F(t={\cal N})=1-S(t={\cal N})=1-Q(t={\cal N})\simeq1$ [see 
Eqs. (\ref{Eq1}), (\ref{Eq2}), (\ref{Eq8}), and (\ref{Eq12})]. 

From the analytically derived functional relation (\ref{Eq16}) one deduces the physically intriguing fact that if the 
time-dependent one-operation success 
probability $s(t)$ of exhausted Sisyphus workers approaches zero asymptotically faster than $t^{-1/{\cal N}}$ 
[that is, if $t\cdot Q(t)\to0$ for $t\to\infty$], then 
the failure probability function of the system is characterized by a non-vanishing asymptotic value, $F(t\to\infty)\neq0$. 
In this case the success probability $S[t;s(t),{\cal N}]$ of the Sisyphus climbers 
to reach their target approaches an asymptotic late-time value which is {\it less} than $1$:
\begin{equation}\label{Eq17}
S[t\to\infty;s(t),{\cal N}]<1\ \ \ \ \text{for}\ \ \ \ t\cdot Q(t\to\infty)\to0\  .
\end{equation}
That is, if $s(t\to\infty)$ approaches zero faster than $t^{-1/{\cal N}}$, then some of the Sisyphus workers 
never complete their repetitive task successfully. 

\section{Time-dependent performance decrement in the scientific literature}
As emphasized above, in many real-life situations the Sisyphus workers 
may gradually become exhausted, in which case the time-dependent probability $1-s(t)$ of a tired climber 
to stumble and fall back to the bottom of the ladder may increase over time. 
In particular, there is empirical evidence in the scientific literature (see \cite{Cap2} and references therein) that 
in some repetitive and monotonous tasks the performance abilities of individuals 
deteriorate exponentially as the time into task increases. 
Considering exhausted Sisyphus workers with exponentially decreasing one-operation 
success probabilities of the form
\begin{equation}\label{Eq18}
s(t)=\alpha\cdot e^{-\beta t}\  ,
\end{equation}
where $\alpha$ and $\beta$ are model-dependent positive constants, 
one finds from Eqs. (\ref{Eq8}), (\ref{Eq16}), and (\ref{Eq18}) that the 
failure probability function of the exhausted Sisyphus random climbers is given by the 
time-dependent functional expression
\begin{equation}\label{Eq19}
F(t)=\exp\Big[
{{{\alpha}^{\cal N}e^{{{{\cal N}({\cal N}+1)}\over{2}}\beta}}\over{{\cal N}\beta}}\cdot
\Big(e^{-{\cal N}\beta t}-e^{-{\cal N}^2\beta}\Big)\Big]\  .
\end{equation} 

From the analytically derived expression (\ref{Eq19}) one deduces that the 
failure probability function approaches the constant finite value
\begin{equation}\label{Eq20}
F(t\to\infty)=\exp\Big[-
{{{\alpha}^{\cal N}e^{-{{{\cal N}({\cal N}-1)}\over{2}}\beta}}\over{{\cal N}\beta}}\Big]>0\  
\end{equation} 
at asymptotically late times, which implies that 
some of the exhausted Sisyphus workers whose performance abilities deteriorate exponentially over time 
never complete their task successfully. 
As expected, the asymptotic failure function (\ref{Eq20}) 
is a monotonically decreasing function of the success pre-factor $\alpha$ and a 
monotonically increasing function of the exponent $\beta$ [see the expression (\ref{Eq18}) 
for the one-operation success probability]. 

Interestingly, one finds that the phase space of the system has a richer structure in situations in which 
the one-operation performance ability of an 
exhausted Sisyphus worker deteriorates over time as an inverse power \cite{Sisy} of the number of jumps (repetitive operations) that 
have already been done. In particular, if the performance ability of the worker remains constant until a critical time $t_0$ and then 
deteriorates as an inverse power of time, 
\begin{equation}\label{Eq21}
s(t)=
\begin{cases}
s_0& \ \ \text{for}\ \ \ \ \ t\leq t_0\  ;
\\ s_0\cdot\big({{t_0}\over{t}}\big)^{\beta} & \ \ \text{for}\ \ \ \ \ t\geq t_0\  ,
\end{cases}
\end{equation}
then one obtains from Eqs. (\ref{Eq8}), (\ref{Eq16}), and (\ref{Eq21}) the functional expressions 
\begin{equation}\label{Eq22}
F(t)\simeq e^{-s^{\cal N}_0(t_0-{\cal N})}\cdot
\begin{cases}
\exp\Big[-\Theta(t-t_0)
{{s^{\cal N}_0t^{\beta{\cal N}}_0}\over{1-\beta{\cal N}}}
\big(t^{1-\beta{\cal N}}-t^{1-\beta{\cal N}}_0\big)\Big]& 
\ \ \text{for}\ \ \ \ \ \beta {\cal N}\neq1\  ;
\\ \Big[\Theta(t-t_0)\cdot\big({{t_0}\over{t}}\big)^{s^{\cal N}_0 t_0}+1-\Theta(t-t_0)\Big]& 
\ \ \text{for}\ \ \ \ \ \beta {\cal N}=1\  
\end{cases}
\end{equation}
for the time-dependent failure probability function of the exhausted workers in the $t_0\gg{\cal N}$ regime, 
where $\Theta(x)$ is the Heaviside step function \cite{NoteHeav}. 

From Eq. (\ref{Eq22}) one deduces that, depending on the magnitude of the dimensionless product $\beta{\cal N}$, 
the failure probability function of the Sisyphus workers can have three qualitatively different 
asymptotic behaviors:
\begin{equation}\label{Eq23}
F(t\to\infty)=e^{-s^{\cal N}_0(t_0-{\cal N})}\cdot
\begin{cases}
\exp\Big({{s^{\cal N}_0 t_0}\over{1-\beta{\cal N}}}\Big)\cdot
\exp\Big(-{{s^{\cal N}_0t^{\beta{\cal N}}_0}\over{1-\beta{\cal N}}}\cdot t^{1-\beta{\cal N}}\Big)& 
\ \ \text{for}\ \ \ \ \ \beta {\cal N}<1\  ;
\\ \big({{t_0}\over{t}}\big)^{s^{\cal N}_0 t_0}& 
\ \ \text{for}\ \ \ \ \ \beta {\cal N}=1\  ;
\\ \exp\Big(-{{s^{\cal N}_0 t_0}\over{\beta{\cal N}-1}}\Big)& 
\ \ \text{for}\ \ \ \ \ \beta {\cal N}>1\  .
\end{cases}
\end{equation}
As expected, one finds that the failure probability function (\ref{Eq23}) is a monotonically decreasing function 
of the success parameter $s_0$ and a 
monotonically increasing function of the exponent $\beta$ in the one-operation success 
probability function (\ref{Eq21}). In particular, the failure function (\ref{Eq22}) 
approaches an asymptotic finite value in the $\beta {\cal N}>1$ regime, 
implying that some of the exhausted Sisyphus workers never complete their task successfully.

\section{Summary and Discussion}
Strong qualitative evidence that has been mounting in the scientific 
literature \cite{Cap1,Cap2,Cap3,Cap4,Cap5,Mesh1,Mesh2,Mesh3,NW1,NW2,NW3} suggests that the 
performance abilities of most people in repetitive monotonous tasks gradually decline as the time-on-task increases. 
In the present work we have provided, 
using a time-dependent Sisyphus random climb model, a quantitative description of this interesting human behavior. 

The stochastic model includes a restart mechanism in which at each repetitive operation the Sisyphus worker (the climber) 
has a time-dependent probability $1-s(t)$ to make a mistake, 
resulting in the need to restart the Sisyphean climbing process again. 
In particular, the empirical evidence \cite{Cap1,Cap2,Cap3,Cap4,Cap5,Mesh1,Mesh2,Mesh3,NW1,NW2,NW3} suggests that 
the stumble probability $1-s(t)$ of a tired worker in a single repetitive operation may 
increase over time. 
Using analytical techniques, we have studied the time-dependent functional behavior of the success probability 
function $S[t;s(t),{\cal N}]$ of the system, a fundamental quantity which describes the fraction of Sisyphus workers who 
have completed their task at time $t$ (or before) by making ${\cal N}$ successful repetitive operations 
in a row without a single stumble in between.

We have revealed the fact that a non-trivial temporal behavior of 
the one-operation success probability function $s(t)$ of realistic exhausted workers 
may have a dramatic effect on the functional behavior of the success probability 
function $S[t;s(t),{\cal N}]$ that characterizes the dynamics of the Sisyphus climbing model (\ref{Eq1}). 
In particular, using analytical techniques, we have proved that the success probability 
function $S[t;s(t),{\cal N}]$ 
of exhausted Sisyphus workers whose one-operation success probability functions approach 
zero asymptotically in time is
given by the remarkably compact asymptotic functional expression [see Eqs. (\ref{Eq4}), (\ref{Eq8}), and (\ref{Eq16})]
\begin{equation}\label{Eq24}
S(t\to\infty)=1-F(t\to\infty)=1-F_0\cdot\exp\Big[-\int_{{\cal N}}^{t} Q(t')dt'\Big]\ \ \ \ \text{for}\ \ \ \ s(t\to\infty)\to0\  .
\end{equation}
Taking cognizance of the fact that tired Sisyphus workers in repetitive monotonous tasks 
are characterized by the asymptotic property $Q(t\to\infty)\to0$, 
one learns from the analytically derived expression (\ref{Eq24}) that these workers are characterized 
by failure probability functions that asymptotically approach a constant finite (non-zero) value or 
decay asymptotically to zero slower than an exponent. 

In particular, from the analytically derived expression (\ref{Eq24}) one deduces the physically intriguing fact that exhausted Sisyphus climbers who are characterized by 
the inverse power law functional behavior $s(t)\sim t^{-1/{\cal N}}$ of their time-dependent 
one-operation success probability mark the boundary between 
Sisyphus workers whose success functions $S[t;s(t),{\cal N}]$ approach $1$ asymptotically in time 
(implying that all the workers eventually complete their task) and 
Sisyphus workers whose success functions approach an asymptotic value 
which is less than $1$ (in which case some of the exhausted Sisyphus climbers never 
complete their task successfully). 

\bigskip
\noindent
{\bf ACKNOWLEDGMENTS}
\bigskip

This research is supported by the Carmel Science Foundation. I would
like to thank Yael Oren, Arbel M. Ongo, Ayelet B. Lata, and Alona B.
Tea for helpful discussions.



\begin{thebibliography}{99}
\bibitem{Notetng} Here one should use the relations $s(t-{\cal N}\cdot\Delta t-\Delta t)\equiv0$ and 
$N_{\text{tot}}(t-{\cal N}\cdot\Delta t-\Delta t)\equiv1$ for $t-{\cal N}\cdot\Delta t-\Delta t<0$.
\bibitem{Notetnnn} Note that the master equation (\ref{Eq11}) is valid in the $t\geq {\cal N}\cdot\Delta t$ regime. 
The failure probability function of the system is characterized by the trivial relation $F\equiv1$ in the 
$t<{\cal N}\cdot\Delta t$ regime [see Eqs. (\ref{Eq2}) and (\ref{Eq5})].
\bibitem{NoteqQ} Note that the strong inequality (\ref{Eq12}) also implies the relation $Q(t\geq{\cal N})\ll1$ [see Eq. (\ref{Eq8})].
\bibitem{Notecl} Here we have used the strong inequality $[{\cal N}\cdot (dF/dt)]/F\ll 1$, which is valid 
in the regime ${\cal N}Q\ll1$ [see Eq. (\ref{Eq8}), (\ref{Eq12}), and (\ref{Eq15})].
\end{thebibliography}
\end{document}